\documentclass[twocolumn,showpacs,preprintnumbers,amsmath,amssymb,10pt,fleqn]{revtex4}

\usepackage[usenames]{color}
\usepackage{colortbl}
\usepackage{ upgreek }

\usepackage{tree-dvips}
\usepackage{graphicx}
\usepackage{dcolumn}
\usepackage{bm}
\usepackage{hyperref}
\usepackage{euscript}
\usepackage{comment}
\input epsf

\begin{document}

\title{Observational Sequences of Randall-Sundrum Black Holes}

\author{Kristina A. Rannu$\dag$, Nikita S. Erin$\ddag$, Stanislav O. Alexeyev$\dag$}

\affiliation{\vskip 3pt
$\dag$Sternberg Astronomical Institute, Lomonosov Moscow State
University, Universitetsky Prospekt, 13, Moscow 119991, Russia}

\affiliation{\vskip 3pt
$\ddag$Department of Astrophysics and Stellar Astronomy, Physics
Faculty, Lomonosov Moscow State University, Vorobievi Gory, 1/2,
Moscow 119991, Russia} 

\date{\today}

\begin{abstract}

We consider the spherically symmetric black hole solution of the Randall-Sundrum (RS) model. This asymptotically-Schwarzschild solution was found by Abdolrahimi et al. (ACPY solution) \cite{a13}. We investigate the specific properties of the accretion disk around ACPY black hole having astrophysical size and mass. The energy flux and temperature distribution are obtained and compared with the Schwarzschild ones. We also present the spectral energy distribution of the accretion disk around RS black holes, when the transfer function is taken into account. Thus we confirm a full agreement of the accretion rate in RS model with the GR predictions. Using recent data on the high-mass X-ray binary Cygnus X-1 and on the Galactic centre black hole candidate Sgr A* we model the accretion in the RS model.  

\end{abstract}

(setq x-select-enable-clipboard t)

\maketitle

\section{Introduction}\label{S:intro} 

Over the years various observational surveys such as the Automatic Plate Measuring (APM) galaxy survey \cite{apm96}, Sloan Digital Sky Survey (SDSS) \cite{sdss09}, etc. were carried out for studying galaxies and clusters. These observations demonstrated that almost all the active galactic nuclei or black hole candidates are surrounded by the gas clouds together with associated accretion disks. The sizes of these accretion disks vary from $0.1$ to several hundreds parsecs \cite{up95}. The most powerful evidence for super massive black holes existence appeares from the Very Long Baseline Interferometry (VLBI). It is imaging molecular H$_2$O masers in the active galaxy NGC 4258 \cite{mm95}. This imaging was produced by Doppler shift measurements
assuming Keplerian motion of the masering source. These results allowed a quite accurate estimation of the central mass: a $3.6 \times 10^7 M/M_{\odot}$ super massive dark object within $0.13$ parsecs. Hence the important astrophysical information can be obtained from the observation of gas streams motion in the gravitational field of compact objects.

Various independent high-precision observational data sets confirm (with startling evidence) that the Universe is undergoing a phase of accelerated expansion \cite{g01}. Several ideas ranging from dark energy models to extended theories of gravity (ETGs) \cite{cl11, mgc12, nr14} were proposed to explain this phenomenon. Specifically ETGs assume that at large scales GR breaks down and the gravitational field is described by a more generic action.

First consequences of the attempts to solve the hierarchy problem were the particle physics modifications around the TeV scale. Therefore the form of gravitational interactions at distances shorter than a millimeter \cite{hw96} is allowed to be more  complicated than usual Newtonian one. In particular such modification could be achieved by introducing of the extra dimensions. In older developments of string theory they were compact \cite{ss74}. Randall and Sundrum presented a new static classical 5D solution of Einstein equations and showed that extra dimensions do not necessary need to be compact.

Two possible applications of this solution were proposed \cite{rs99, rs100}. In the first case (which we refer to as RSI) there are two
branes with non-vanishing tensions placed at the orbifold fixed points \cite{rs99}. Our Universe lives on the brane with negative tension (``the visible brane''). All the matter describing by three fundamental interactions (strong, weak and electro-magnetic) is localized on this brane. However gravity is able to propagate, in addition, along the 5th dimension into the bulk. The exponential warp factor appearing in the RSI model leads to a new natural solution of the hierarchy problem \cite{rs99}. In the second (RSII) proposal \cite{rs100} the visible brane has the positive tension while the second brane is moved to the infinity. Such topology provides an exciting example of a non-compact extra direction, which nevertheless correctly reproduces Newton's law on the visible brane. However the hierarchy problem is not completely solved in this particular case. RSII model needs an appropriately tuned value of the tension on the brane. Such delicate adjustment is equivalent to the cosmological constant problem. The brane tension and bulk cosmological constant define the shape of the bound state graviton mode. In addition to the bound state one, there is a continuum of the Kaluza-Klein modes. These states have weak coupling to low-energy states on the brane, but are essential for the consistency of the full theory of gravity and would couple strongly to Planck-energy brane processes \cite{rs100}.

The mass accretion on rotating black holes in stationary axisymmetric spacetimes in the frames of GR was studied in details long ago \cite{bpt72, nt73, ss73, ss76, ss86}. The radiation emitted by the accretion disk surface is studied under the assumption of
thermodynamical equilibrium of the disk. That is why black body radiation is a good approximation in this case \cite{pt74}. So there is a possibility to test the RS model using astrophysical observations of the emission spectra from accretion disks. The purpose of the present paper is to study the properties of a thin accretion disk around an RSII black hole and carry out an analysis of the radiation emerging from the surface of the disk.

The paper is organized as follows. In Sec. \ref{S:disk} we review the formalism and the physical properties of the thin disk
accretion onto compact objects for stationary axisymmetric spacetimes. In Sec. \ref{S:RSBHs} we analyze the basic properties of
matter forming a thin accretion disk around large black holes in the RSII model. We discuss and conclude our results in
Sec. \ref{S:concl}. Throughout this work we use a system of units so that $c = G = \hbar = k_B = 1$, where $k_B$ is Boltzmann's constants. 

\section{Accretion disk}\label{S:disk}

The description of accretion process onto black holes was developed in details by Bardeen and Press \cite{bpt72}, Novikov and Thorne \cite{nt73}, Shakura and Sunyaev \cite{ss73, ss76, ss86}, Page and Thorne \cite{pt74}. The particles constituting the accretion disk are assumed to move in circular orbits in the frames of GR. The physical properties and the electromagnetic radiation characteristics of these particles are determined by the geometry of the spacetime around the compact object. For a stationary and axially symmetric geometry the metric is given in a general form as
\begin{gather} \label{eq:metric}
  ds^2 = g_{tt} dt^2 + 2 g_{t\phi} dt d\phi + g_{rr} dr^2 + g_{\theta\theta} d\theta^2 + g_{\phi\phi} d\phi^2. 
\end{gather}
In the present work we consider a thin disk accretion, therefore the equatorial approximation is used. The metric components $g_{tt}$, $g_{t\phi}$, $g_{rr}$, $g_{\theta\theta}$ and $g_{\phi\phi}$ depend on the radial coordinate $r$ only, and $|\theta - \pi/2| \ll 1$.

To compute the relevant physical quantities of thin accretion disks we determine the radial dependence of the specific angular velocity $\Omega$, the specific energy $\tilde{E}$ and the specific angular momentum $\tilde{L}$ of particles \cite{hk08, hk09}:
\begin{gather} \label{eq:emv}
  \begin{split}
    \tilde{E} &= - \ \cfrac{g_{tt} + g_{t\phi} \Omega}{\sqrt{- g_{tt} - 2 g_{t\phi} \Omega - g_{\phi\phi} \Omega^2}}, \\
    \tilde{L} &= \cfrac{g_{t\phi} + g_{\phi\phi} \Omega}{\sqrt{-g_{tt} - 2 g_{t\phi} \Omega - g_{\phi\phi} \Omega^2}}, \\
    \Omega &= \cfrac{d\phi}{dt} = \cfrac{- g_{t\phi,r} + \sqrt{(g_{t\phi,r})^2 - g_{tt,r} g_{\phi\phi,r}}}{g_{\phi\phi,r}}.
  \end{split}
\end{gather}
The marginally stable orbit around the central object is given by
\begin{gather} \label{eq:rms}
  0 = \tilde{E}^2 g_{\phi\phi,rr} + 2 \tilde{E} \tilde{L} g_{t\varphi,rr} + \tilde{L}^2 g_{tt,rr} - (g_{t\phi}^2 - g_{tt} g_{\phi\phi})_{,rr},
\end{gather}
where $g_{t\phi}^2 - g_{tt} g_{\phi\phi}$ (appearing as a cofactor in the metric determinant) never vanishes. Inserting \eqref{eq:emv} into \eqref{eq:rms} and solving the obtained equation for $r$, we find the radii of the marginally stable orbits, once the metric coefficients $g_{tt}$, $g_{t\phi}$ and $g_{\phi\phi}$ are explicitly given.

For a thin accretion disk its vertical size is assumed negligible compared to its horizontal extension. So the disk height $H$,
defined by the maximum of half disk thickness, is always much smaller than the characteristic radius $r$ of the disk, $H \ll r$. The thin disk has an inner edge at the marginally stable orbit of the compact object potential, and the accreting plasma has a Keplerian motion in higher orbits.

In steady-state accretion disk models the mass accretion rate $\dot{M_0}$ is assumed to be a constant in time. This quantity
measures the rate at which the rest mass of the particles flow inward through the disk. The radiation flux $F$ emitted by the surface of the accretion disk can be derived from the conservation equations for the mass, energy and angular momentum respectively. Then the radiant energy $F(r)$ over the disk is expressed in terms of the specific energy, angular momentum and the angular velocity of the particles orbiting in the disk \cite{nt73, pt74, hk08, hk09}:
\begin{gather} \label{eq:f}
  F(r) = - \ \frac{\dot{M_0}}{4 \pi \sqrt{-g}} \
  \cfrac{\Omega_{,r}}{(\tilde{E} - \Omega \tilde{L})^2} \int \limits_{r_{ms}}^{r} (\tilde{E} - \Omega \tilde{L}) \ \tilde{L}_{,r} \ dr,
\end{gather}
where $r_{ms}$ is the marginally stable orbit obtained from \eqref{eq:rms}.

The other characteristic of the accretion process is conversion efficiency. This value indicates how fast the central object converts
the rest mass into the outgoing radiation. The efficiency is defined as the ratio of the rate of the radiation energy of photons, escaping from the disk surface to infinity, and the rate at which mass-energy is transported to the central compact object. Both values are measured at the infinity \cite{nt73, pt74}. If all the emitted photons can escape to infinity, the efficiency is given in terms of the specific energy measured at the marginally stable orbit $r_{ms}$:
\begin{gather} \label{eq:eff}
  \varepsilon = 1 - \tilde{E}_{ms}.
\end{gather}
For Schwarzschild black holes the efficiency $\varepsilon$ is about $6 \%$, whether the photon capture by the black hole is considered or not.

The accreting matter in the steady-state thin disk is supposed to be in thermodynamical equilibrium. Therefore the radiation emitted by a disk surface could be considered as a perfect black body radiation. Thus the energy flux is given by $F(r) = \sigma T^4(r)$, where $\sigma$ is the Stefan-Boltzmann constant. The observed luminosity $L(\nu)$ has a redshifted black body spectrum \cite{t02}: 
\begin{gather} \label{eq:l}
  L(\nu) = 4\pi d^2 I(\nu) = \cfrac{8}{\pi} \cos{\gamma} \int   \limits_{r_{i}}^{r_{f}} \int \limits_{0}^{2\pi}
  \cfrac{{\nu^3_e} r d \phi dr}{\exp\left( h \nu_e / T \right) - 1},
\end{gather}
where $d$ is the distance to the source, $I(\nu)$ is the thermal energy flux radiated by the disk (Planck distribution function),
$\gamma$ is the disk inclination angle, $r_i$ and $r_f$ indicate the position of the inner and outer edge of the disk
respectively. Following the work \cite{hk09}, we expect the flux over the disk surface to vanish at $r \rightarrow \infty$ for any kind of general relativistic compact object geometry. Therefore we take $r_i = r_{ms}$ and $r_f \rightarrow \infty$. The emitted frequency is given by $\nu_e = \nu(1 + z)$, and the transfer function (redshift factor) can be written as \cite{hk09}
\begin{gather} \label{eq:tf}
  1 + z = \cfrac{1 + \Omega r \sin{\varphi} \sin{\gamma}}{\sqrt{ - g_{tt} - 2 \Omega g_{t\varphi} - \Omega^2 g_{\varphi\varphi}}},
\end{gather}
where the light bending is neglected \cite{l79}.

In the next sections we will calculate the energy flux, temperature and luminosity of the accretion disk around Schwarzschild and RSII black holes. We adopt the following values for the two fundamental mass classes: $M = 14.8 M_{\odot}$, $\dot{M_0} = 7.484 \times 10^{-7}M/M_{\odot}$, which are the best estimates available for the well-known stellar-mass black hole Cygnus-X1 \cite{o11, m11}, and $M = 4 \times 10^{6}M_{\odot}$, $\dot{M_0} = 3 \times 10^{-5}M/M_{\odot}$ for the Galactic centre black hole candidate Sgr A* \cite{bn06, df13, fm13}.

\section{Randall-Sundrum black holes}\label{S:RSBHs}

\subsection{ACPY solution}

Several black hole solutions in RS model were presented in \cite{rsbh1}. However for a long time it was argued that stable static
black hole solutions were applicable for RSII. Some obtained solutions had a horizon radius not greater than the AdS length $\ell$
\cite{rsbh2}. Some authors \cite{rsbh3} proposed that stable RSII black holes could not exist at all. Obviously, such an argument serves an extremely strong evidence against the viability of the RSII model.

Nevertheless new solutions for large stable black holes were found recently \cite{fw11, a13}. The one obtained by Abdolrahimi, Catto\"en, Page and Yaghoobpour-Tari (ACPY solution) \cite{a13} is especially interesting. It represents a spherically symmetric static and asymptotically Schwarzschild black hole solution. ACPY solution was found using an explicit approximate metric for the black holes on the brane. Abdolrahimi et al. have constructed an infinite mass axisymmetric 5D black hole solution of the bulk Einstein equation. This black hole solution with infinite mass was used to find the metric of a large Randall-Sundrum II black hole on the brane \cite{a13}. We refer the reader to \cite{a13} for details and present the final metric given by
\begin{gather}\label{eq:met}
  \begin{split}
    ds^2 &= \gamma_{\mu\nu} dx^{\mu}dx^{\nu} = - \left( 1 - \cfrac{2M}{r} \right) dt^2 + \\
    &+ \left [1 - \cfrac{1}{ - {\Lambda_5} r^2} \ \cfrac{r - 2M}{r - 1.5M} \left( \EuScript{F} - r \ \cfrac{d \EuScript{F}}{dr}
      \right) \right] \times \\
    &\times \left( 1 - \cfrac{2M}{r} \right)^{-1} \! \! \! \! \! \! dr^2 + \left[ r^2 +
      \cfrac{\EuScript{F}}{ - \Lambda_5} \right] d{\Omega}^2,
  \end{split}
\end{gather}
where $\Lambda_5$ is the 5D cosmological constant characterizing the vacuum's model qualities and $\EuScript{F}$ is a numerically found 11th-order polinom given by
\begin{gather}\label{eq:pol}
  \begin{split}
    \EuScript{F} \approx 1 &- 1.1241 \left( \cfrac{2M}{r} \right) +     1.956 \left( \cfrac{2M}{r} \right)^2 - \\
    &- 9.961 \left( \cfrac{2M}{r} \right)^3 + \ \dots \ + 2.900 \left( \cfrac{2M}{r} \right)^{11},
  \end{split}
\end{gather}
with the normalized mean-square error $J_{\EuScript{F}} \approx 0.0000572$.

In our previous work \cite{ar15} we considered the weak field limit for the RSII black hole solutions by Figueras and Wiseman \cite{fw11} and by Abdolrahimi et al. \cite{a13}. We showed the good agreement with GR in PPN limit for both solutions. Now we compare the accretion properties of the ACPY solution with the standard Schwarzschild metric case.

\subsection{Integration Constants and Conversion Efficiency}

Inserting the metric components of \eqref{eq:met} into the expressions \eqref{eq:emv} of the specific energy, specific angular momentum and the angular velocity, we obtain
\begin{gather}\label{eq:res}
  \begin{split}
    \tilde{E} &= \cfrac{1 - \cfrac{2M}{r}}{\sqrt{1-\cfrac{2M}{r} - \left(r^2 + \cfrac{\EuScript{F}}{-\Lambda_5} \right) {\Omega}^2 }}, \\
    \tilde{L} &= \cfrac{\left( r^2 - \cfrac{\EuScript{F}}{ - \Lambda_5} \right) \Omega }{\sqrt{1 - \cfrac{2M}{r} -
        \left(r^2 + \cfrac{\EuScript{F}}{ - \Lambda_5} \right) {\Omega}^2 }}, \\
    \Omega & = \sqrt{ \cfrac{2M/r^2}{ 2r + \cfrac{ d\EuScript{F}/dr}{- \Lambda_5}}}. 
  \end{split}
\end{gather}
As \eqref{eq:res} shows, the motion constants for the particles orbiting in the equatorial plane depend on radial coordinate $r$
only. The non-vanishing function $\EuScript{F}$ gives a negligible contribution to the volume element
\begin{gather}\label{eq:grs}
  \begin{split}
    \sqrt{- g_{RS}} &= \left(r^2 + \cfrac{\EuScript{F}}{- {\Lambda_5}} \right ) \times \\
    &\times \sqrt{1 - \cfrac{1}{- {\Lambda_5} r^2} \ \cfrac{r - 2M}{r - 1.5M} \left(\EuScript{F} - r \cfrac{d\EuScript{F}}{dr} \right)}
  \end{split}
\end{gather}
as compared to the case of the equatorial approximation for Schwarzschild black holes with $\sqrt{-g_{GR}} = r^2$.

\begin{widetext}

\begin{figure}[t]
\begin{center}
  {\includegraphics[height=6cm]{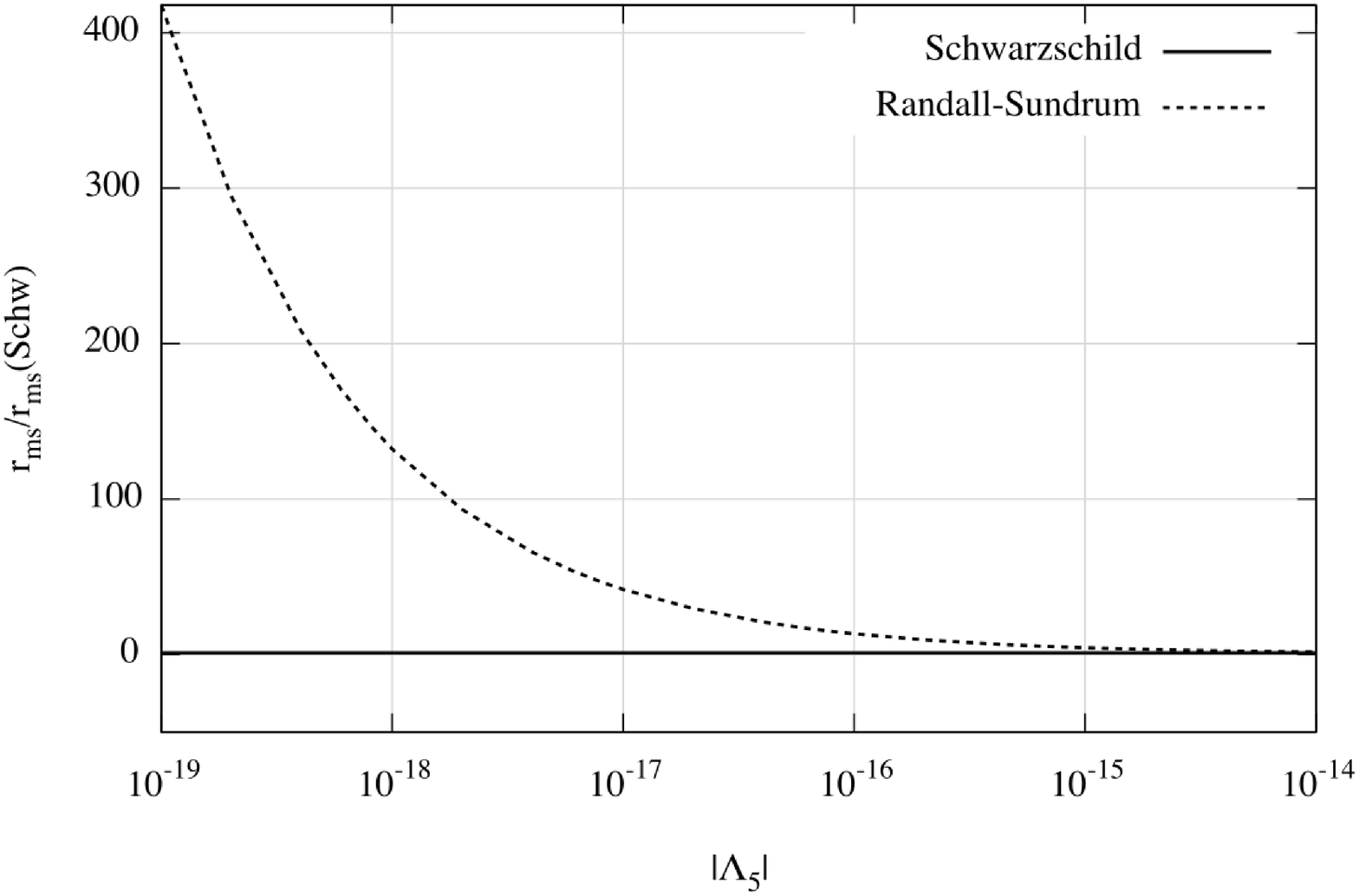}\includegraphics[height=6cm]{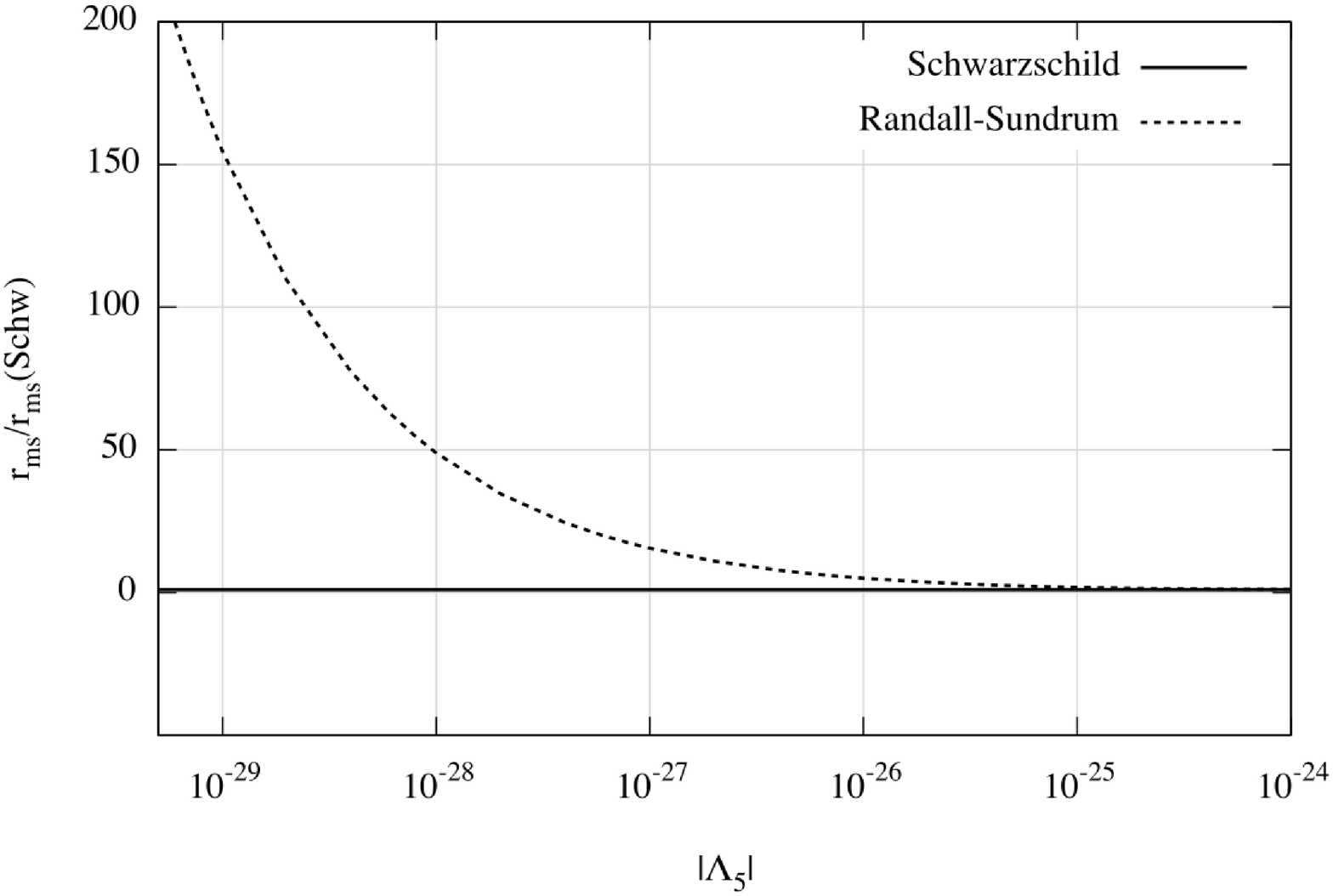}}
\end{center}
\caption{The radii of the marginally stable orbits $r_{ms}$ for the ACPY black holes compared to the Schwarzschild values against the bulk cosmological constant's modulo value $|\Lambda_5|$ for $M = 14.8 M_{\odot}$ (left) and $M = 4 \times 10^6 M_{\odot}$ (right).}
\label{rms}
\end{figure}
\begin{figure}[t]
\begin{center}
  {\includegraphics[height=6.2cm]{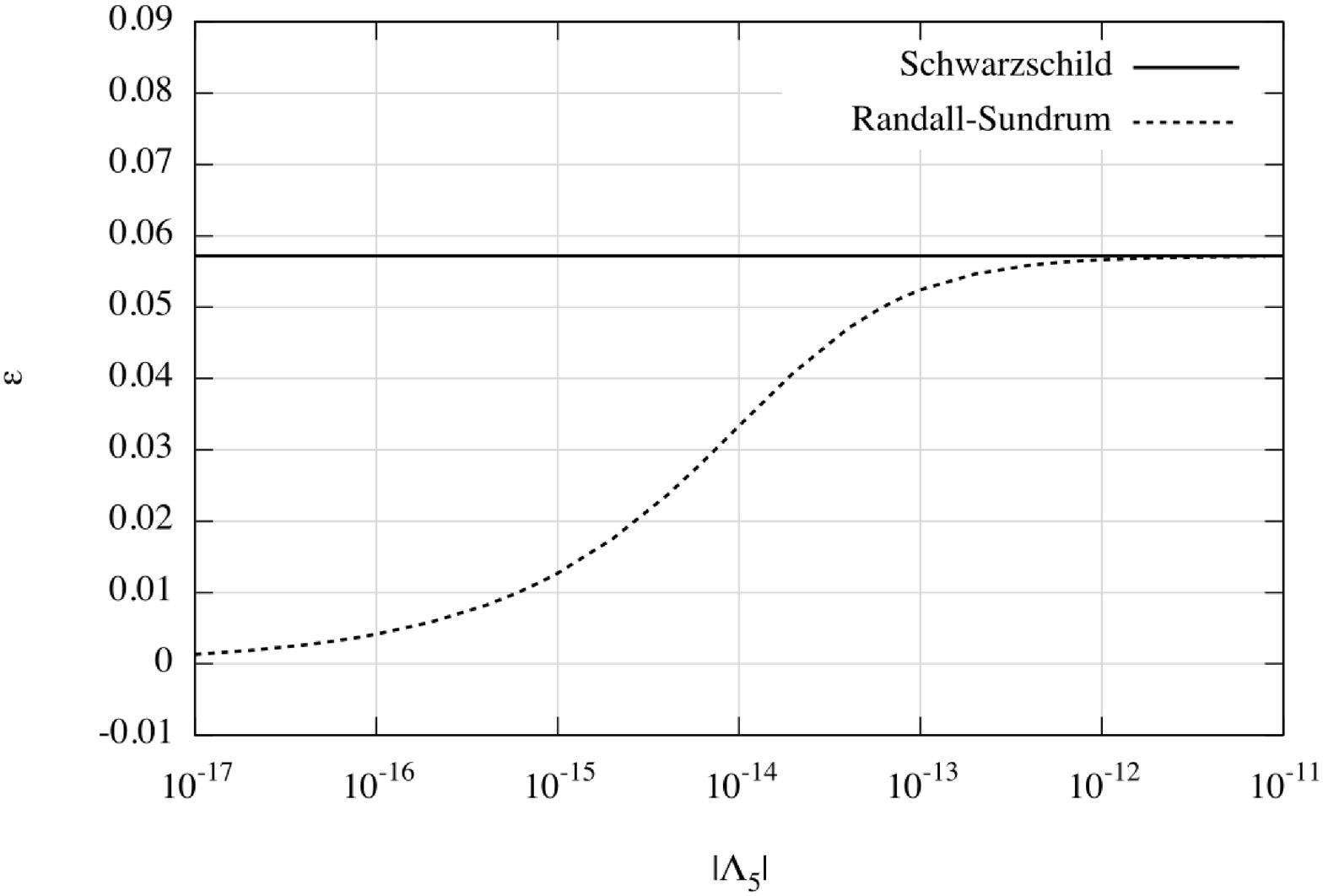}\includegraphics[height=6.2cm]{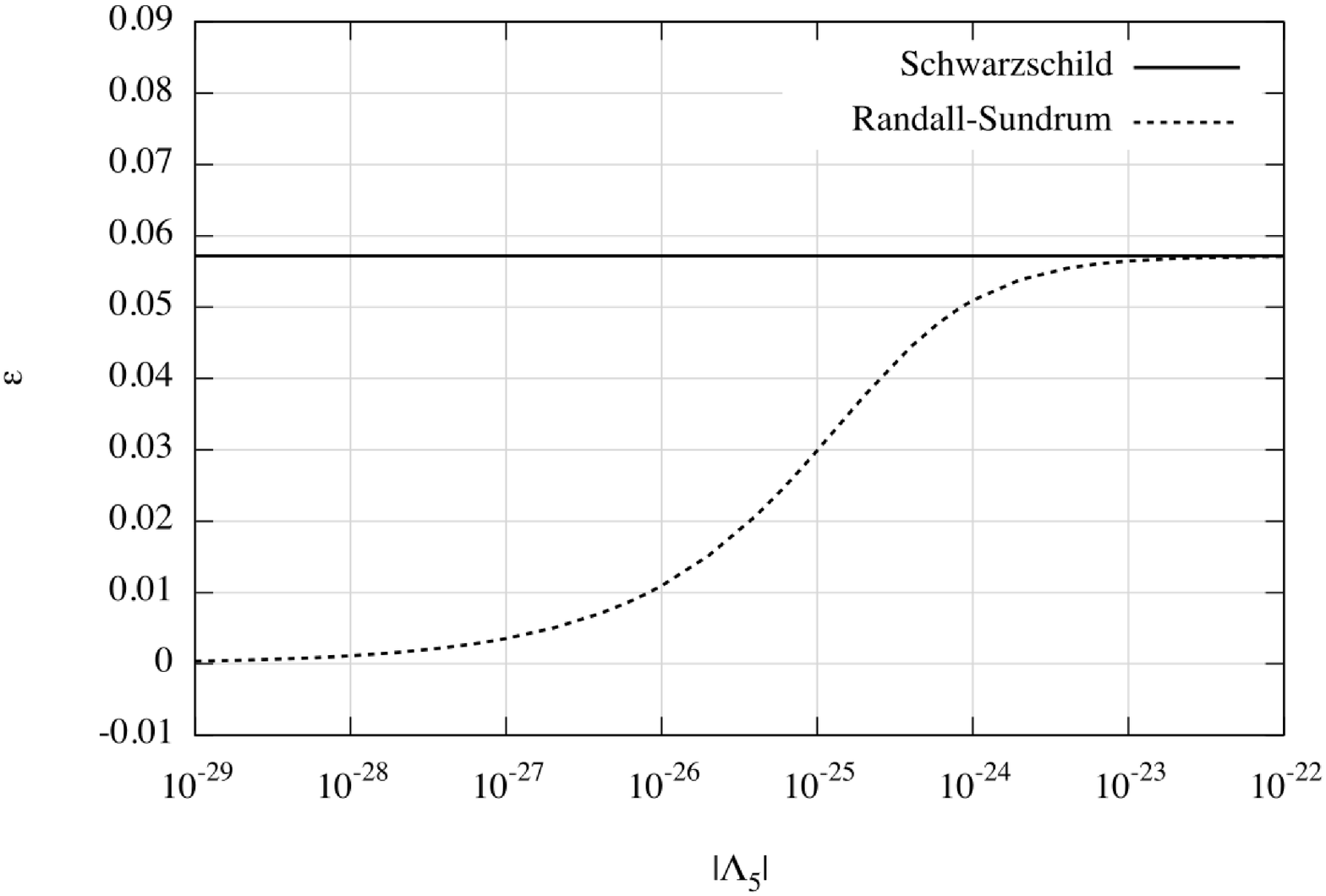}}
\end{center}
\caption{Conversion efficiency $\varepsilon$ for Schwarzschild black holes and the ACPY black holes against the bulk cosmological constant's modulo value $|\Lambda_5|$ in the bulk for $M = 14.8 M_{\odot}$ (left) and $M = 4 \times 10^6 M_{\odot}$ (right).}
\label{eff}
\end{figure}

\end{widetext}

As a result, the properties of particles orbiting in the equatorial plane of Schwarzschild and RSII black holes are essentially the
same. The main difference is the location of the marginally stable orbits (FIG. \ref{rms}), and the accretion efficiency $\varepsilon$, (FIG. \ref{eff}). However these quantities are affected by the bulk cosmological constant $\Lambda_5 \gtrsim -10^{-13}$ for stellar-mass black holes ($M = 14.8 M_{\odot}$) and $\Lambda_5 \gtrsim -10^{-24}$ for supermassive black holes ($M = 4 \times 10^6 M_{\odot}$).

\subsection{Flux, temperature and spectrum of the disk}

In FIG. \ref{f} and FIG. \ref{t} we present the flux and temperature distribution for the Schwarzschild and the ACPY black holes
respectively. The cosmological constant $\Lambda_5$ runs from $-2 \times 10^{-15}$ to $-5 \times 10^{-14}$ a stellar-mass black hole with $M = 14.8 M_{\odot}$ and from $-2 \times 10^{-25}$ to $-5 \times 10^{-26}$ to for a supermassive black hole with $M = 4 \times 10^6 M_{\odot}$.

The plots in FIG. \ref{f} show that the energy flux profiles of the disks within the RSII model deviate from the Schwarzschild black hole case if $\Lambda_5 \to 0$, as expected. For the largest values of $\Lambda_5$ the inner edge of the accretion disk is located at higher radius than the inner edge of the disk around Schwarzschild black hole (FIG. \ref{rms}). As the quantities $\tilde{E}$, $\tilde{L}$ and $\Omega$ in the flux integral (\ref{eq:f}) for the RSII model are still close to those for the Schwarzschild case, for the higher boundary $r_{ms}(ACPY) > r_{ms}(GR)$, the integral gives lower flux values. Thus the maximum of the integrated flux is smaller in the RSII model and decreases further while increasing the bulk cosmological
\begin{widetext}

\begin{figure}[t]
\begin{center}
  {\includegraphics[height=6.4cm]{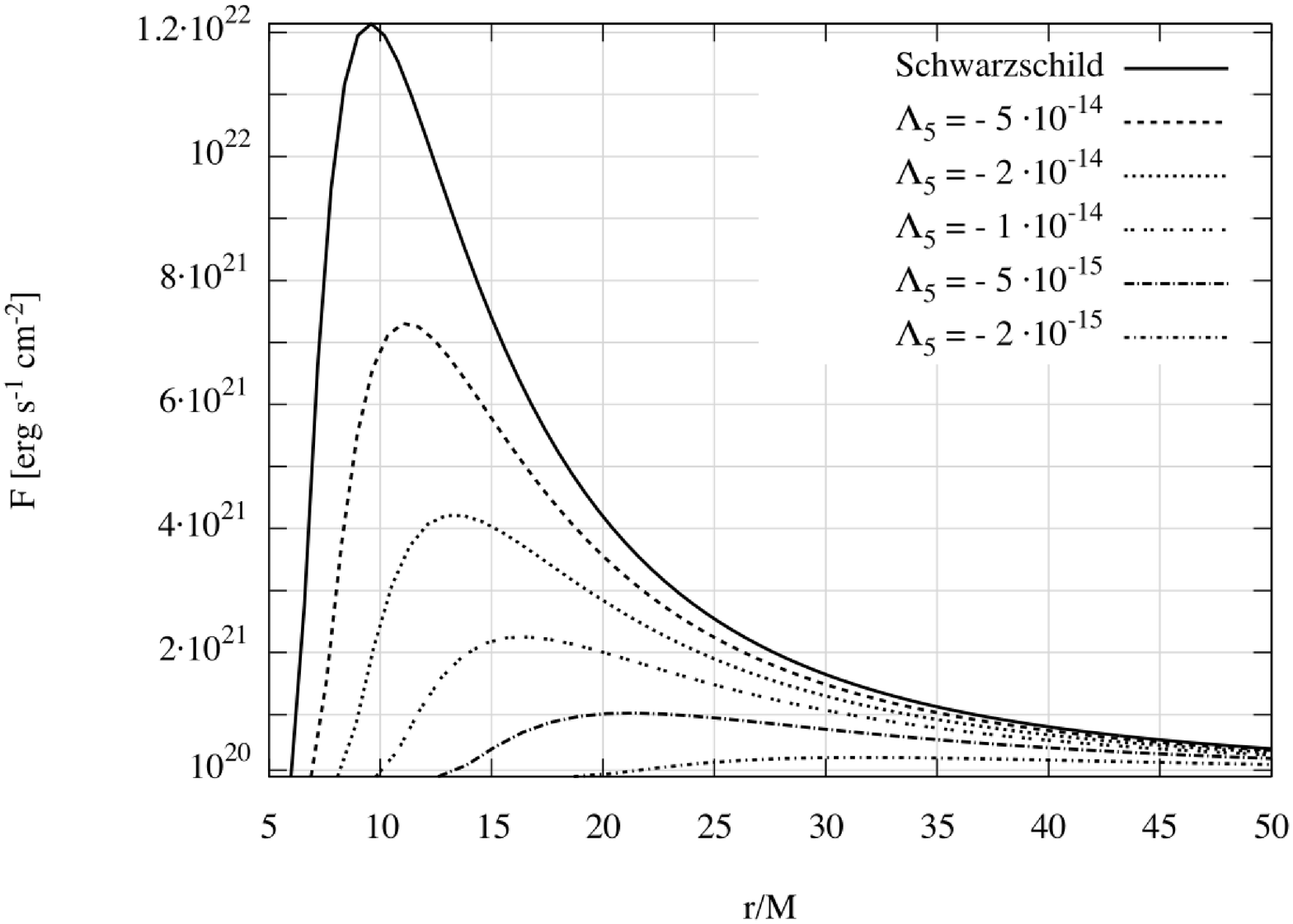}\includegraphics[height=6.4cm]{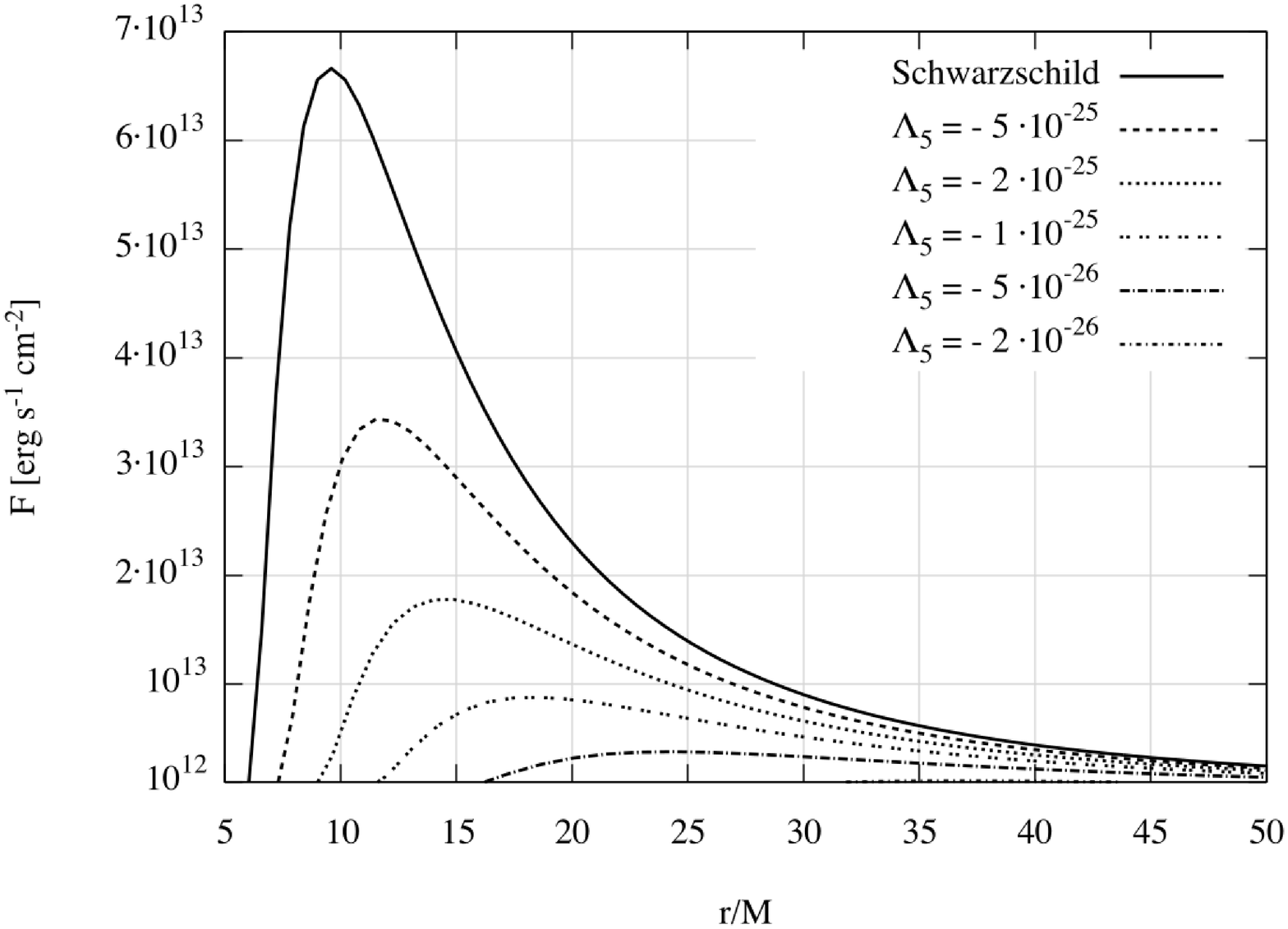}}
\end{center}
\caption{The energy flux $F [\mbox{erg}/(\mbox{s} \ \mbox{cm}^2)]$ distribution over the disks for Schwarzschild black holes and the ACPY black holes against cosmological constant in the bulk $\Lambda_5$ for $M = 14.8 M_{\odot}$ (left) and $M = 4 \times 10^6 M_{\odot}$ (right).}
\label{f}
\end{figure}
\begin{figure}[t]
\begin{center}
  {\includegraphics[height=6.4cm]{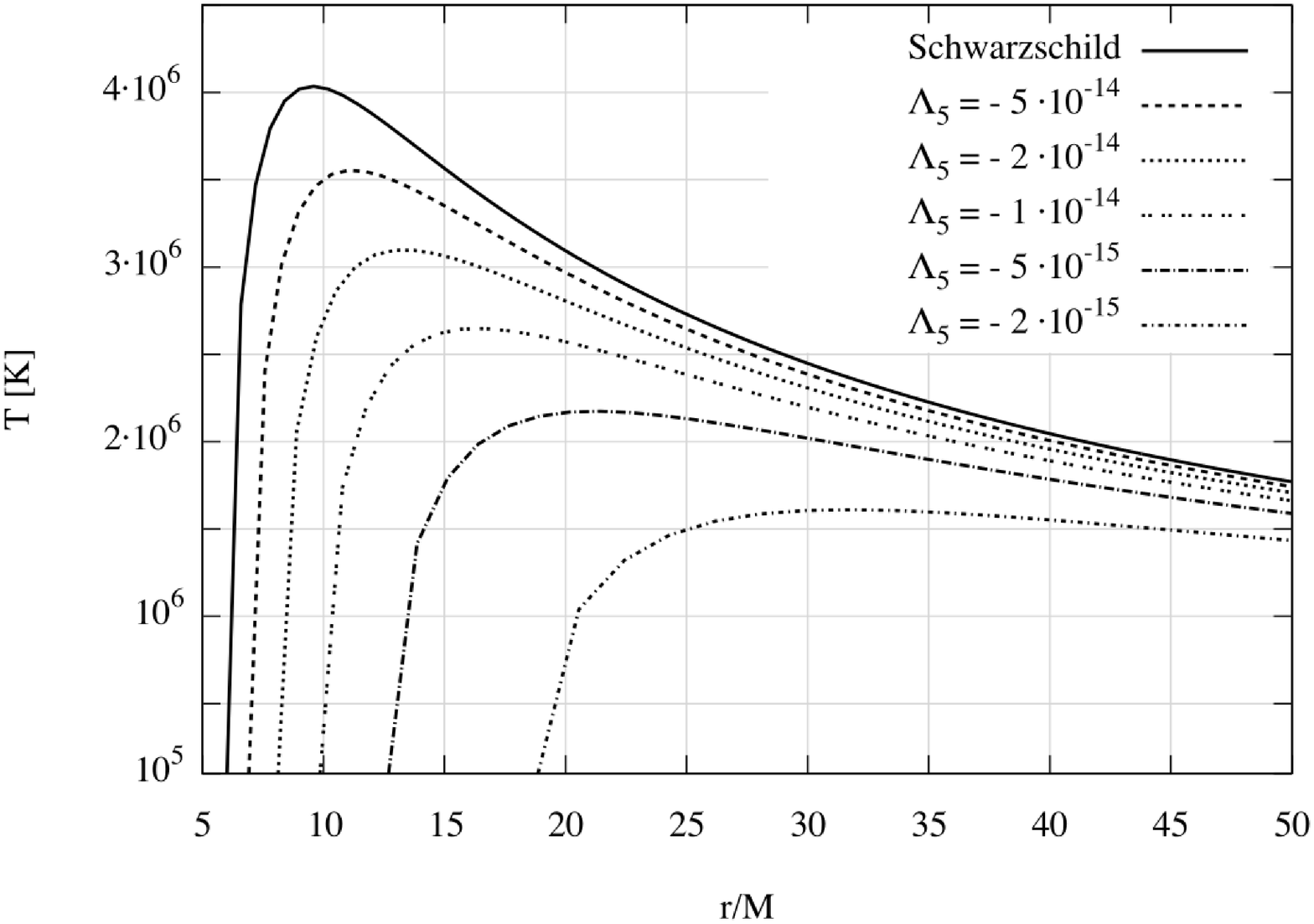}\includegraphics[height=6.4cm]{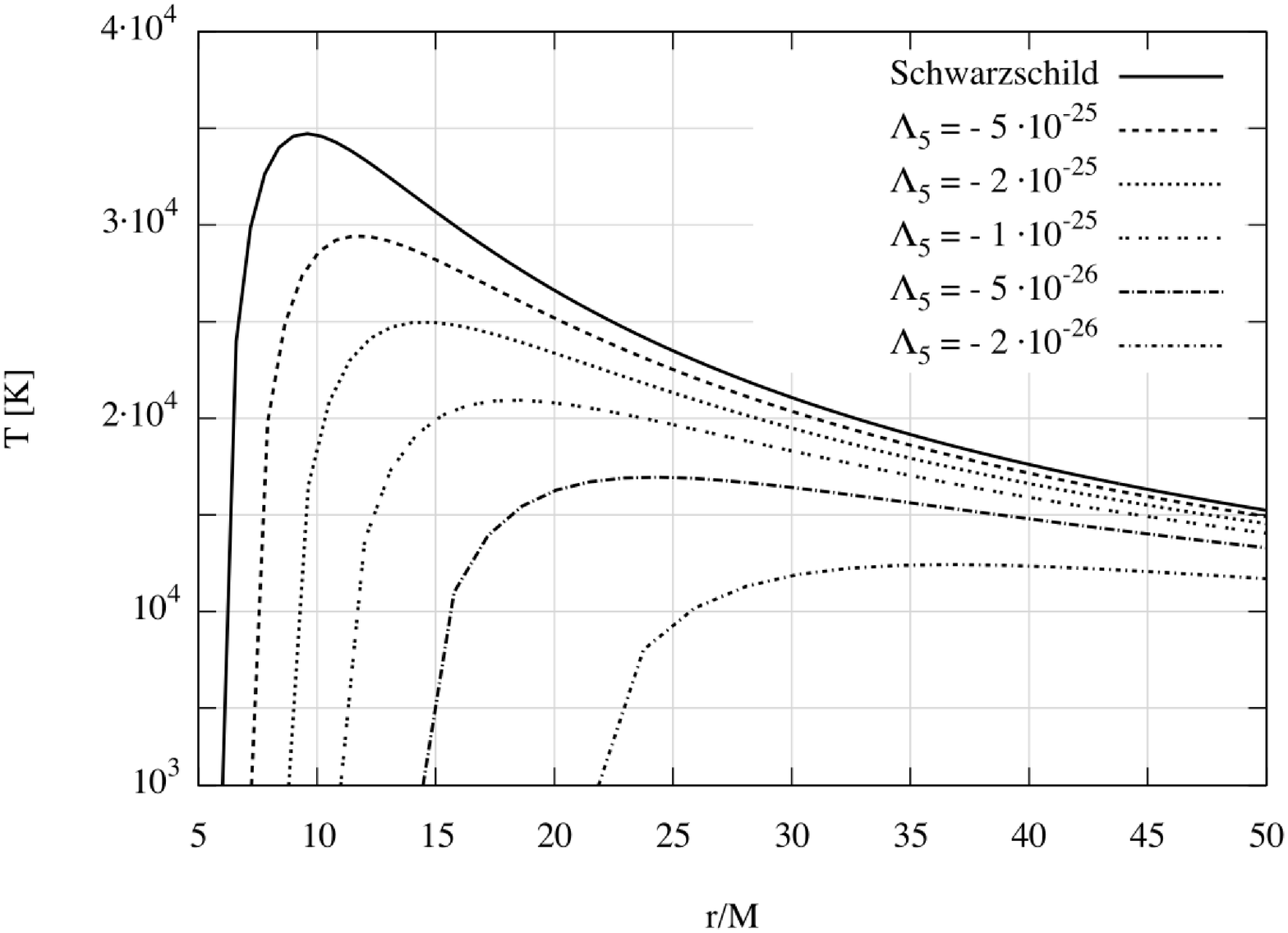}}
\end{center}
\caption{The temperature $T [K]$ distribution over the disks for Schwarzschild black holes and the ACPY black holes against cosmological constant in the bulk $\Lambda_5$ for $M = 14.8 M_{\odot}$ (left) and $M = 4 \times 10^6 M_{\odot}$ (right).}
\label{t}
\end{figure}

\end{widetext}
constant. The difference of the ACPY flux maximum from the Schwarzschild achieves $\sim 10^3$ times for stellar black holes and $\sim 10^2$ for supermassive ones.

Similar features can be found while plotting the temperature profiles of the disk (FIG. \ref{t}). As the accretion disk moves off from the ACPY black hole while increasing the bulk cosmological value $\Lambda_5$, the temperature gets lower in $\sim 2.5$ times for stellar black holes and $\sim 3$ for supermassive ones. However the differences here are less than for the energy flux, since the temperature is proportional to $F^{1/4}$.

\begin{widetext}

\begin{figure}[t]
\begin{center}
  {\includegraphics[height=6.5cm]{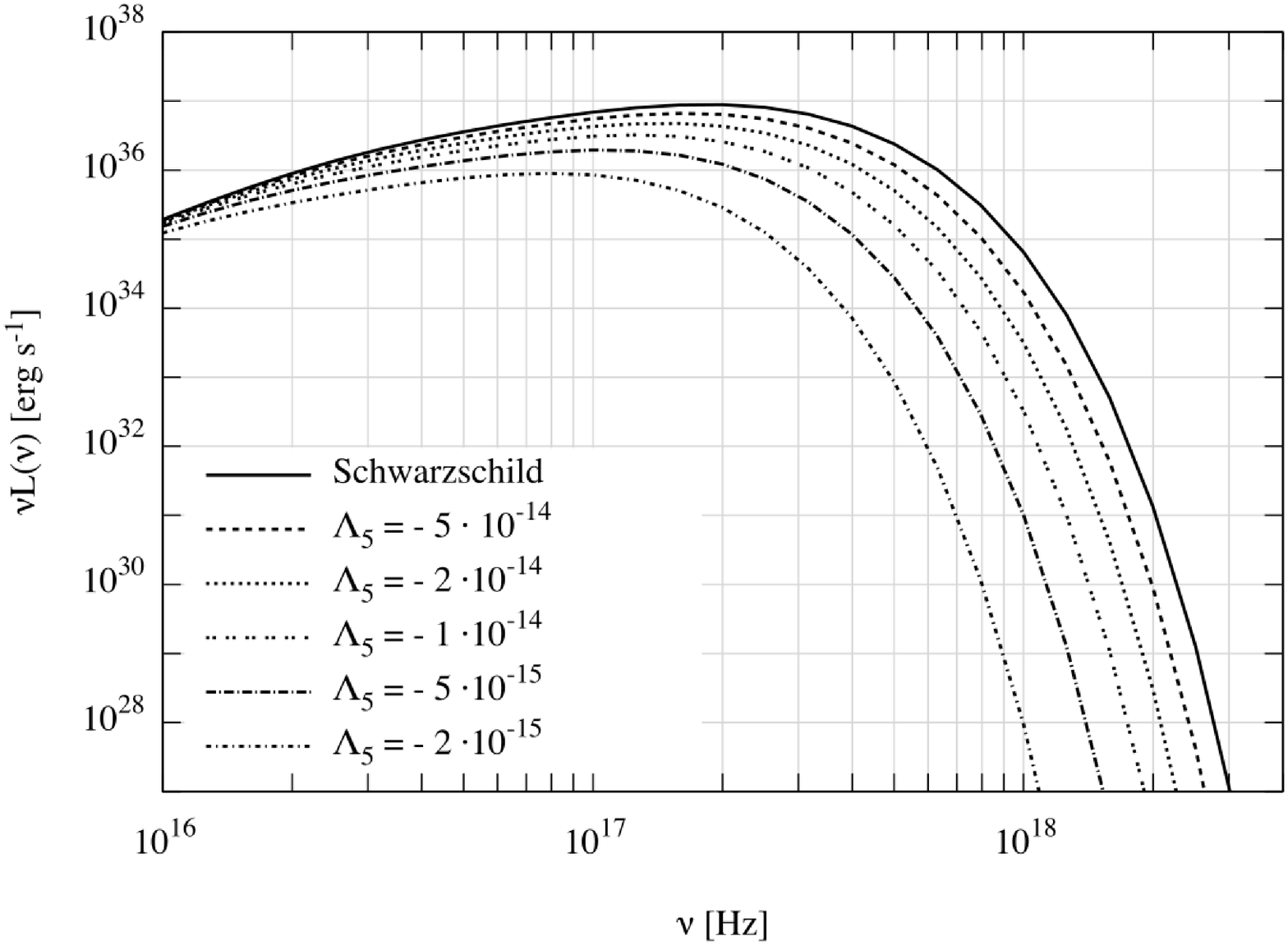}\includegraphics[height=6.5cm]{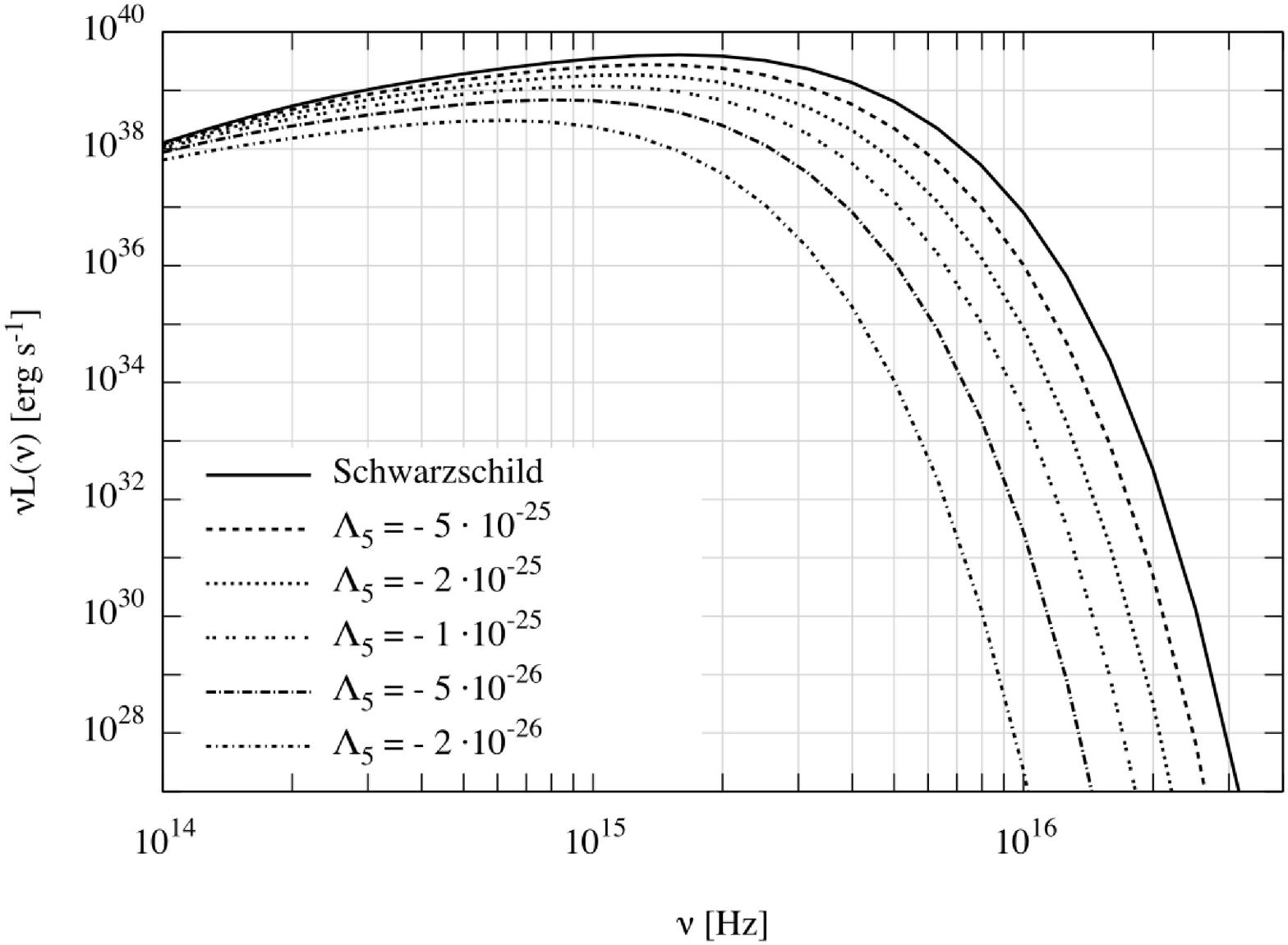}}
\end{center}
\caption{The accretion disks spectra $\nu L(\nu) [\mbox{erg/s}]$, $\nu   [\mbox{Hz}]$ for Schwarzschild black holes and the ACPY black holes against cosmological constant in the bulk $\Lambda_5$ for $M = 14.8 M_{\odot}$ (left) and $M = 4 \times 10^6 M_{\odot}$ (right).}
\label{sp}
\end{figure}

\end{widetext}

The variation of the emission spectrum for different values of the bulk cosmological constant is presented in FIG. \ref{sp}. The plots
show that while $\Lambda_5$ in the RS-gravity increases, the cut-off frequency of the accretion disk spectrum decreases from the
Schwarzschild value by $1/5$ of order approximately in both cases. Therefore they reach the values about the $1.1 \times 10^{18}$ Hz for a black hole with $M = 14.8 M_{\odot}$ type and $10^{16}$ Hz for a black hole with $M = 4 \times 10^6 M_{\odot}$ instead of $1.2 \times 10^{18}$ Hz and $1.21 \times 10^{16}$ Hz respectively.

However all these differences occur at very specific values of bulk cosmological constant $\Lambda_5$. Generally the radiation of the accretion disk around ACPY black holes produces the spectra similar to the standard Schwarzschild case. Therefore the ACPY solution has a full agreement with GR in the matter of accretion.

\section{Conclusions}\label{S:concl}

In the present work we consider the matter forming a thin accretion disk in stationary axisymmetric black hole spacetime within the RSII model. We took into account the black holes of two mass ranges. Cgn X-1 with $M = 14.8 M_{\odot}$ was taken as the example of the black hole with stellar mass. Sgr A* with $M = 4 \times 10^{6} M_{\odot}$ was used as the example of the supermassive black hole.

The main physical parameters of the disk are the energy flux distribution, the temperature distribution and emission spectrum. These quantites and the additional ones such as the radii of the marginally stable orbits and the conversion efficiency, were
explicitly obtained for the astrophysical-range black holes solution in RSII by Abdolrahimi, Catto\"en, Page and Yaghoobpour-Tari with different values of the bulk cosmological constant $\Lambda_5$. This solution appears to be asymptotically Schwarzschild up to very small modulo values of the bulk cosmological constant $\Lambda_5$. The notable differences from the Schwarzschild case in the accretion disk properties appear at least for $\Lambda > -10^{-12}$. However the ACPY black hole solution was constructed for a bulk cosmological constant $\Lambda_5 = - 6$ in the units of the curvature length scale of the negatively curved five-dimensional AdS spacetime (AdS length). That means that in the range of definition of the ACPY solution shows a very good consistency both with the GR predictions and the results of observations.

Abdolrahimi et al. have made an important step proving the existence of black holes of astrophysical scales within the RSII model
\cite{a13}. Our results presented in this paper confirm the reliability of their solution. Together with our previous conclusions
on the weak-field (PPN) limit \cite{ar15} they are a strong argument in favor of the ACPY solution and therefore in favor of the RSII model as a whole.

\section*{Acknowledgments}

Authors would like to thank professors N.I. Shakura, K.S. Thorne and T.S. Harko for very useful discussions and great help on performing of this work. This work was partially supported by individual grant from Dmitry Zimin Foundation ``Dynasty'' (S.A.).

\end{document}